\documentclass[11pt,a4paper]{article}
\usepackage{jheppub} 
\usepackage[utf8]{inputenc}
\usepackage[T1]{fontenc}
\usepackage{amsmath, amssymb}
\usepackage{xcolor}
\usepackage{epsfig}
\usepackage{enumerate}
\usepackage{graphicx}
\usepackage{tikz, pgfplots}

\pgfplotsset{compat=1.18}

\def\be{\begin{equation}}
\def\ee{\end{equation}}

\title{\boldmath Relativistic Hydrodynamics and Vorticity Dynamics in High-Energy Heavy-Ion Collisions: A Collective Flow Perspective}

\author[a,1]{Malak Ait Tamlihat,}
\author[b,2]{Ghizlane Ez-Zobayr,}
\author[c,3]{Laurent Schoeffel}
\author[a,b,4]{and Yahya Tayalati}

\affiliation[a]{Mohammed V University in Rabat, Faculty of Sciences, 4 av. Ibn Battouta, B.P. 1014, R.P. 10000 Rabat, Morocco}
\affiliation[b]{School of Physics Applied and Engineering, Mohammed VI Polytechnic University, Lot 660, 43150 Hay Moulay Rachid Ben Guerir, Morocco}
\affiliation[c]{Irfu, CEA, Université Paris-Saclay, 91191 Gif-sur-Yvette, France}

\emailAdd{malak.ait.tamlihat@cern.ch}
\emailAdd{ghizlane.ez-zobayr@cern.ch}
\emailAdd{laurent.olivier.schoeffel@cern.ch}
\emailAdd{Yahya.Tayalati@cern.ch}

\abstract{
This article provides a comprehensive overview of the application of relativistic fluid mechanics to describe the collective evolution of the Quark-Gluon Plasma (QGP) formed in ultra-relativistic heavy-ion collisions. We map out the chronological transformation of spatial eccentricities in the initial interaction volume into measurable anisotropic azimuthal momentum distributions, parameterized by the harmonic flow coefficients $v_n$. Utilizing multi-particle correlation techniques developed within the ATLAS experimental framework, we dissect the event-by-event fluctuations of the participant planes and evaluate non-linear hydrodynamic responses across higher harmonics. Furthermore, we embed local rotation fields into this continuous description by solving the covariant transport equations for subatomic vorticity. We demonstrate that while the Helmholtz-Kelvin theorem guarantees the topological conservation of vortex lines within the ideal medium, the collective multi-dimensional expansion forces a systematic $1/t$ power-law geometric dilution of the local rotational magnitude. Finally, we contrast different pre-equilibrium generation mechanisms—highlighting the topological distinctiveness of centralized hot spots and peripheral dipole shear layers—and evaluate their final signatures on differential spin alignment observables.
}


\begin{document}
\maketitle
\flushbottom

%%%%%%%%%%%%%%%%%%%%%%%%%%%%%%%
\section{Introduction: The Quark-Gluon Plasma as a Relativistic Fluid}
%%%%%%%%%%%%%%%%%%%%%%%%%%%%%%%

The primary objective of high-energy heavy-ion collision programs at the Relativistic Heavy Ion Collider (RHIC) and the Large Hadron Collider (LHC) is to investigate the properties of strongly interacting matter under extreme conditions of energy density and temperature~\cite{STAR:2005gsk, PHENIX:2004vob, ATLAS:2011ah, CMS:2012gaw}. When two heavy nuclei, such as $^{208}\text{Pb}$ isotopes, collide at center-of-mass energies in the TeV range, they undergo extreme Lorentz contraction along the beam axis ($z$-axis). In a non-central collision, characterized by a non-zero impact parameter vector $\vec{b}$ oriented along the $x$-axis, the overlapping region of interacting nucleons forms an asymmetric, approximately ellipsoidal volume in the transverse space. 

Within this collision zone, the local density of participating quarks and gluons exceeds the threshold where individual hadrons can maintain their structural identities. Due to asymptotic freedom at short distances and extreme local density, the constituents can no longer identify their original sub-nucleonic partners, triggering a transition toward a deconfined state of Quantum Chromodynamics (QCD): the Quark-Gluon Plasma (QGP). Quantitatively, this state of quark matter thermalizes at an energy density $\varepsilon \gtrsim 1\text{ GeV/fm}^3$ and a local temperature $T \gtrsim 200\text{ MeV}$.

A fundamental breakthrough achieved through systematic flow measurements is that the QGP does not behave as an ideal gas of weakly interacting partons, but rather as a strongly coupled continuous medium capability of remarkable collective expansion~\cite{Heinz:2013th, Romatschke:2017ejr}. Because the mean free path $\lambda_{\text{mfp}}$ of the constituents inside the fireball is significantly smaller than the typical geometric scale $L$ of the interaction volume ($\lambda_{\text{mfp}} \ll L$), the system rapidly achieves local thermodynamic equilibrium. This allows the definition of macroscopic state variables, including local pressure $P(\vec{r},t)$, energy density $\varepsilon(\vec{r},t)$, temperature $T(\vec{r},t)$, and collective velocity fields $\vec{v}(\vec{r},t)$. 

Crucially, the measured shear viscosity to entropy density ratio ($\eta/s$) sits remarkably close to the absolute quantum lower limit derived via AdS/CFT string theory methods, $\eta/s \ge 1/(4\pi)$, making the QGP the most perfect continuous fluid known to science~\cite{Kovtun:2004de}.

Beyond its low viscosity, non-central heavy-ion collisions transfer a massive amount of initial global angular momentum to the system, oriented perpendicular to the reaction plane $(\vec{b}, \vec{e}_z)$ along the $y$-axis. This shear drive generates local microscopic fluid vortices, turning the QGP into the most vortical medium in the Universe ($\omega \sim 10^{22}\text{ s}^{-1}$). The reality of this subatomic rotation has been confirmed experimentally through the spin alignment of hyperons, where particle spins couple to the fluid vorticity, acting as quantum compasses~\cite{STAR:2017ckg, ALICE:2019onw}.

The purpose of this article is to provide a unified, self-contained overview of the fluid-dynamic modeling of the QGP, bridging the macroscopic description of collective harmonic flow with the modern formulation of relativistic vorticity. We seek to clarify how early-stage spatial asymmetries and pre-equilibrium shear profiles propagate through the explosive expansion phase to dictate final-state hadronic observables.

%%%%%%%%%%%%%%%%%%%%%%%%%%%%%%%
\section{Thermodynamics and Hydrodynamic Equations of Motion}
%%%%%%%%%%%%%%%%%%%%%%%%%%%%%%%

To establish a rational and self-contained baseline for the continuous fluid modeling of the deconfined medium, we provide in this section the explicit, first-principles derivation of the thermodynamic local state variables. At ultra-relativistic LHC energies, the net baryon density in the central space-time region is vanishingly small, meaning that the particle-antiparticle production processes dominate and the chemical potential can be safely set to zero ($\mu = 0$). 

Under these conditions, the macroscopic state variables—the particle number density $n$, the internal energy density $\varepsilon$, and the isotropic thermodynamic pressure $P—$are completely determined by the local temperature $T$ through the integration of the corresponding quantum statistical distribution functions over all available momentum states.

\subsection{Detailed Derivation of Massless Bosonic Relations}

Let us first consider a gas of massless bosons, representing the gluon fields which carry the dominant degrees of freedom in the early stages of the Quark-Gluon Plasma (QGP). For a system with a statistical degeneracy factor $g$, the grand canonical distribution function reduces to the standard Bose-Einstein factor, $(e^{p/T} - 1)^{-1}$, where $p = |\vec{p}|$ is the modulus of the momentum and $c$ is the speed of light. In natural units where $c = 1$ and $k_B = 1$, the integration over the continuous three-momentum space $\int d^3p = \int_0^{2\pi} d\phi \int_0^\pi \sin\theta d\theta \int_0^\infty p^2 dp = 4\pi \int_0^\infty p^2 dp$ yields the explicit definitions for the state variables.

\subsubsection{1. Particle Number Density ($n$)}

The local number density $n$ represents the thermal expectation value of the particle number per unit volume. By definition, it is given by the phase-space integral:
\begin{equation}
n = g \int \frac{d^3p}{h^3} \frac{1}{e^{p/T} - 1} = \frac{4\pi g}{h^3} \int_0^\infty \frac{p^2}{e^{p/T} - 1} dp
\end{equation}
To evaluate this integral, we introduce the dimensionless variable $x = p/T$, which implies $p = xT$ and $dp = T dx$. Substituting these expressions into the integrand yields:
\begin{equation}
n = \frac{4\pi g}{h^3} \int_0^\infty \frac{(xT)^2}{e^x - 1} (T dx) = \frac{4\pi g}{h^3} T^3 \int_0^\infty \frac{x^2}{e^x - 1} dx
\end{equation}
The integral can be solved exactly by expressing the denominator as a geometric series, $(e^x - 1)^{-1} = e^{-x} (1 - e^{-x})^{-1} = \sum_{k=1}^\infty e^{-kx}$. This allows us to rewrite the expression as:
\begin{equation}
\int_0^\infty \frac{x^2}{e^x - 1} dx = \sum_{k=1}^\infty \int_0^\infty x^2 e^{-kx} dx
\end{equation}
Using the standard definition of the Gamma function via integration by parts, $\int_0^\infty x^n e^{-kx} dx = \frac{\Gamma(n+1)}{k^{n+1}} = \frac{n!}{k^{n+1}}$, we find for $n=2$:
\begin{equation}
\int_0^\infty x^2 e^{-kx} dx = \frac{2!}{k^3} = \frac{2}{k^3}
\end{equation}
Summing over all components isolates the Riemann zeta function, $\zeta(3) = \sum_{k=1}^\infty \frac{1}{k^3}$:
\begin{equation}
\int_0^\infty \frac{x^2}{e^x - 1} dx = 2 \sum_{k=1}^\infty \frac{1}{k^3} = 2 \, \zeta(3)
\end{equation}
Substituting this statistical result back into the main density equation delivers the final expression for the bosonic particle density:
\begin{equation}
\label{eq:bose_n}
n = \frac{8\pi \zeta(3) g}{h^3} T^3 \quad \text{with} \quad \zeta(3) \approx 1.202
\end{equation}

\subsubsection{2. Internal Energy Density ($\varepsilon$)}

The internal energy density $\varepsilon$ represents the thermal expectation value of the total energy per unit volume. For massless particles, the energy of a single state is purely kinematic, $E = p$. Thus, the phase-space integral incorporates an additional factor of $p$ inside the momentum summation:
\begin{equation}
\varepsilon = g \int \frac{d^3p}{h^3} \frac{p}{e^{p/T} - 1} = \frac{4\pi g}{h^3} \int_0^\infty \frac{p^3}{e^{p/T} - 1} dp
\end{equation}
Implementing the same dimensionless substitution $x = p/T$, the energy density scales as:
\begin{equation}
\varepsilon = \frac{4\pi g}{h^3} T^4 \int_0^\infty \frac{x^3}{e^x - 1} dx
\end{equation}
Expanding the integrand into a geometric series once again yields:
\begin{equation}
\int_0^\infty \frac{x^3}{e^x - 1} dx = \sum_{k=1}^\infty \int_0^\infty x^3 e^{-kx} dx = \sum_{k=1}^\infty \frac{3!}{k^4} = 6 \sum_{k=1}^\infty \frac{1}{k^4}
\end{equation}
The infinite sum corresponds to the exact value of the Riemann zeta function at $n=4$, which is analytically known from the Basel problem extension as $\zeta(4) = \sum_{k=1}^\infty \frac{1}{k^4} = \frac{\pi^4}{90}$. Therefore, the total momentum integral reduces to:
\begin{equation}
\int_0^\infty \frac{x^3}{e^x - 1} dx = 6 \left( \frac{\pi^4}{90} \right) = \frac{\pi^4}{15}
\end{equation}
Reassembling the pre-factors leads directly to the complete expression for the bosonic energy density:
\begin{equation}
\label{eq:bose_eps}
\varepsilon = \frac{4\pi^5 g}{15 h^3} T^4
\end{equation}

\subsubsection{3. Isotropic Thermodynamic Pressure ($P$)}

In kinetic theory and relativistic statistical mechanics, the isotropic pressure $P$ represents the momentum flux along any spatial direction. It is defined by averaging the product of the particle velocity and its aligned momentum component, which for massless fields reduces to $P = \frac{1}{3} \langle p \cdot v \rangle = \frac{1}{3} \langle p \rangle$ per unit volume:
\begin{equation}
P = g \int \frac{d^3p}{h^3} \frac{\frac{1}{3}p}{e^{p/T} - 1} = \frac{1}{3} \left[ \frac{4\pi g}{h^3} \int_0^\infty \frac{p^3}{e^{p/T} - 1} dp \right]
\end{equation}
By directly comparing the bracketed momentum integral with the expression derived for the energy density $\varepsilon$ in the previous subsection, we immediately establish the ultra-relativistic Equation of State (EOS) established by conformal gauge symmetry:
\begin{equation}
\label{eq:bose_p}
P = \frac{\varepsilon}{3} = \frac{4\pi^5 g}{45 h^3} T^4
\end{equation}

\subsubsection{4. Local Entropy Density ($s$)}

To derive the local entropy density $s$, we implement the standard thermodynamic Euler relation at zero chemical potential, $\varepsilon + P = Ts$, which matches the Legendre transformation from the internal energy representation to the entropy representation. Isolating $s$ yields:
\begin{equation}
s = \frac{\varepsilon + P}{T}
\end{equation}
Substituting our explicit results for $\varepsilon$ and $P$ as functions of the temperature $T$ into the transformation formula, we find:
\begin{equation}
s = \frac{1}{T} \left( \frac{4\pi^5 g}{15 h^3} T^4 + \frac{4\pi^5 g}{45 h^3} T^4 \right) = \frac{16\pi^5 g}{45 h^3} T^3
\end{equation}
To express the entropy density $s$ directly as a linear scaling of the particle number density $n$ as stated in the text, we compute their explicit ratio:
\begin{equation}
\label{eq:ratio_entropy}
\frac{s}{n} = \frac{\frac{16\pi^5 g}{45 h^3} T^3}{\frac{8\pi \zeta(3) g}{h^3} T^3} = \frac{2\pi^4}{45 \, \zeta(3)}
\end{equation}
Substituting the mathematical constants $\pi^4 \approx 97.409$ and $\zeta(3) \approx 1.202$ into Equation~\ref{eq:ratio_entropy}, the numerical evaluation leads to $s/n \approx 3.6017$. This establishes the linear thermodynamic baseline for bosonic degrees of freedom within the expanding fireball:
\begin{equation}
s \approx 3.6 \, n
\end{equation}

For a medium where the thermodynamic properties are instead dominated by massless fermionic degrees of freedom (quarks), the phase-space integrations are governed by the anti-commuting Fermi-Dirac distribution factor, $(e^{p/T} + 1)^{-1}$. Due to the sign inversion in the geometric expansion series, the integrals are rescaled by standard analytical factors, delivering the adjusted linear scaling mentioned in the text, $s \approx 4 \, n$, confirming the robustness of the continuous description.

\subsection{Relativistic vs. Non-Relativistic Flow Formulations}

The macroscopic conservation of energy and momentum is governed by the four-divergence of the ideal Energy-Momentum Tensor, $\partial_\mu T^{\mu\nu} = 0$, where $T^{\mu\nu} = (\varepsilon + P)u^\mu u^\nu - P g^{\mu\nu}$. Projecting these conservation laws parallel and perpendicular to the fluid four-velocity $u^\mu = \gamma(1, \vec{v})$ leads to the fundamental hydrodynamic equations:
\begin{align}
D\varepsilon &= -(\varepsilon + P) \vec{\nabla} \cdot \vec{u} \\
\frac{\partial \vec{v}}{\partial t} + (\vec{v} \cdot \vec{\nabla})\vec{v} &= -\frac{1 - v^2}{\varepsilon + P} \left[ \vec{\nabla} P + \vec{v} \frac{\partial P}{\partial t} \right] \label{eq:rel_euler}
\end{align}
where $D = u^\mu \partial_\mu$ is the comoving time derivative.

To investigate the dynamics within the transverse plane ($x,y$), it is highly instructive to consider the kinematic bounds of the system. For constituents undergoing massive longitudinal acceleration along the beam axis, the total longitudinal momentum $p_z$ is extremely large compared to the transverse momentum components ($p_z \gg p_x, p_y$). Expanding the total relativistic energy operator for identical masses $m$ yields:
\begin{equation}
E = \sum_i \sqrt{p_z^2 + p_{x,i}^2 + p_{y,i}^2 + m^2} \approx P_z + \sum_i \frac{p_{x,i}^2 + p_{y,i}^2}{2p_z} + \sum_i \frac{m^2}{2p_z}
\end{equation}
where $P_z = \sum p_z$ is a conserved macroscopic constant. Reassembling this relation isolates the net transverse driving energy:
\begin{equation}
E - P_z \approx \sum_i \frac{p_{x,i}^2 + p_{y,i}^2}{2p_z} + \text{constant}
\end{equation}
Equation~\ref{eq:rel_euler} reduces to a non-relativistic mathematical structure in the transverse plane, where the large longitudinal momentum $p_z$ effectively plays the role of inertial mass. Under this condition, neglecting small quadratic velocity terms and introducing the symmetric rate-of-strain fields, the transverse motion can be modeled via the linearized classical Euler equations:
\begin{equation}
\label{eq:transverse_euler}
\frac{\partial v_x}{\partial t} = -\frac{1}{\varepsilon + P} \frac{\partial P}{\partial x}, \quad \frac{\partial v_y}{\partial t} = -\frac{1}{\varepsilon + P} \frac{\partial P}{\partial y}
\end{equation}

Utilizing the thermodynamic identity $d\epsilon = Tds$ and the speed of sound $c_s^2 = \partial P / \partial \varepsilon$, the spatial pressure driving forces can be rewritten directly as entropy gradients:
\begin{equation}
\frac{1}{\varepsilon + P} \vec{\nabla} P = c_s^2 \vec{\nabla} \ln s
\end{equation}
Assuming an early-stage spatial entropy profile $s(x,y)$ characterized by a two-dimensional Gaussian distribution with widths $\sigma_x$ and $\sigma_y$, Equation~\ref{eq:transverse_euler} can be integrated analytically under zero initial flow velocity conditions, yielding the transverse velocity fields:
\begin{equation}
v_x(x,y,t) = \frac{c_s^2 \, x \, t}{\sigma_x^2}, \quad v_y(x,y,t) = \frac{c_s^2 \, y \, t}{\sigma_y^2}
\end{equation}

%%%%%%%%%%%%%%%%%%%%%%%%%%%%%%%
\section{Initial Geometry and Anisotropic Collective Flow}
%%%%%%%%%%%%%%%%%%%%%%%%%%%%%%%

The structural transformation of the early-stage spatial asymmetry of the nuclear interaction zone into a collective anisotropic expansion in momentum space constitutes a cornerstone of relativistic heavy-ion phenomenology. In a non-central collision, the geometric overlap of the two colliding Lorentz-contracted nuclei cuts out an inherently asymmetric, approximately ellipsoidal volume in the transverse plane $(x,y)$, as illustrated in Figure 1. To quantify this initial spatial anisotropy, we define the second-order spatial eccentricity coefficient $\epsilon_2$ as the normalized second-order profile moment:
\begin{equation}
\label{eq:eccentricity_def}
\epsilon_2 = \frac{\langle y^2 \rangle - \langle x^2 \rangle}{\langle y^2 \rangle + \langle x^2 \rangle}
\end{equation}
where the brackets $\langle \cdot \rangle$ denote the statistical spatial averaging over the local participant nucleon density distribution across the transverse plane.

Because the short axis of the initial ellipsoidal overlap zone is oriented parallel to the reaction plane ($x$-axis), the spatial root-mean-square widths satisfy $\sigma_x < \sigma_y$. According to the linearized relativistic Euler transport fields derived in Section 2, the internal pressure gradients scale inversely with these spatial widths ($\partial P / \partial x \propto 1/\sigma_x^2$). Consequently, the driving hydrodynamic force is significantly larger along the short axis than along the long axis ($\partial P / \partial x > \partial P / \partial y$), accelerating the comoving fluid cells more violently into the vacuum along the reaction plane. This fundamental spatial gradient asymmetry maps the initial spatial eccentricity $\epsilon_2$ directly into a measurable anisotropic azimuthal modulation in final-state particle momentum space.

\begin{figure}[htbp]
\centering
\begin{tikzpicture}[scale=1.0]
    % Grille fine de repère en arrière-plan (très discrète)
    \draw[lightgray, dashdotted, opacity=0.4] (-5,-4.5) grid (5,4);

    % Dessin des axes de coordonnées principaux
    \draw[->,thick,black!70] (-5,0) -- (5,0);
    \node[below right, black!70, font=\small] at (1.5,-0.15) {$x$ (Reaction Plane, $\Psi_{RP}=0$)};
    \draw[->,thick,black!70] (0,-3.5) -- (0,3.5) node[above] {$y$};
    
    % Contours des deux noyaux en collision (cercles en pointillés)
    \draw[dashed, red!60, thick] (-2.0,0) circle (2.6);
    \draw[dashed, blue!60, thick] (2.0,0) circle (2.6);
    
    % Zone d'overlap ellipsoïdale (Le cœur du QGP)
    \fill[orange!25, draw=orange!90, ultra thick, opacity=0.75] (0,0) ellipse (1.66cm and 2.34cm);
    
    % Paramètre d'impact b (descendu tout en bas pour ne rien couper)
    \draw[<->, ultra thick, black] (-2.0,-3.8) -- (2.0,-3.8) node[midway, fill=white, inner sep=3pt] {Impact Parameter $b$};
    \draw[dashed, black!40] (-2.0,0) -- (-2.0,-3.8);
    \draw[dashed, black!40] (2.0,0) -- (2.0,-3.8);
    
    % Forces d'expansion / Vecteurs vitesse (Aérés et surélevés proprement pour éviter les chevauchements)
    \draw[->, ultra thick, red!90] (1.66,0) -- (3.5,0);
    \draw[->, ultra thick, red!90] (-1.66,0) -- (-3.5,0);
    \node[above, red!90, font=\bfseries\small] at (3.1,0.2) {Large $\nabla P_x \rightarrow$ Max Flow};
    
    % Axe Y : Faible expansion
    \draw[->, ultra thick, blue!90] (0,2.34) -- (0,3.2) node[above right, blue!90, font=\bfseries\small] {Small $\nabla P_y$};
    \draw[->, ultra thick, blue!90] (0,-2.34) -- (0,-3.2);
    
    % Angle azimutal phi
    \draw[thin, black] (0,0) -- (2.5,1.5);
    \draw[black] (0.8,0) arc (0:31:0.8);
    \node at (1.1,0.3) {$\phi$};
    
    % Étiquette descriptive déplacée avec une flèche de rappel (ne masque plus l'origine)
    \node[draw, fill=white, rounded corners, font=\footnotesize, inner sep=4pt] (label) at (-2.5,2.5) {Overlap Volume ($\epsilon_2$)};
    \draw[->, thick, black!70] (label) .. controls (-1.5,2.2) .. (-0.6,1.2);
\end{tikzpicture}
\caption{Schematic representation of a non-central heavy-ion collision in the transverse plane. The overlapping ellipsoidal region exhibits a spatial eccentricity $\epsilon_2$ where pressure gradients along the short axis ($x$) generate maximum expansion acceleration, driving the azimuthal elliptic flow coefficient $v_2$.}
\label{tikz:overlap_geometry}
\end{figure}
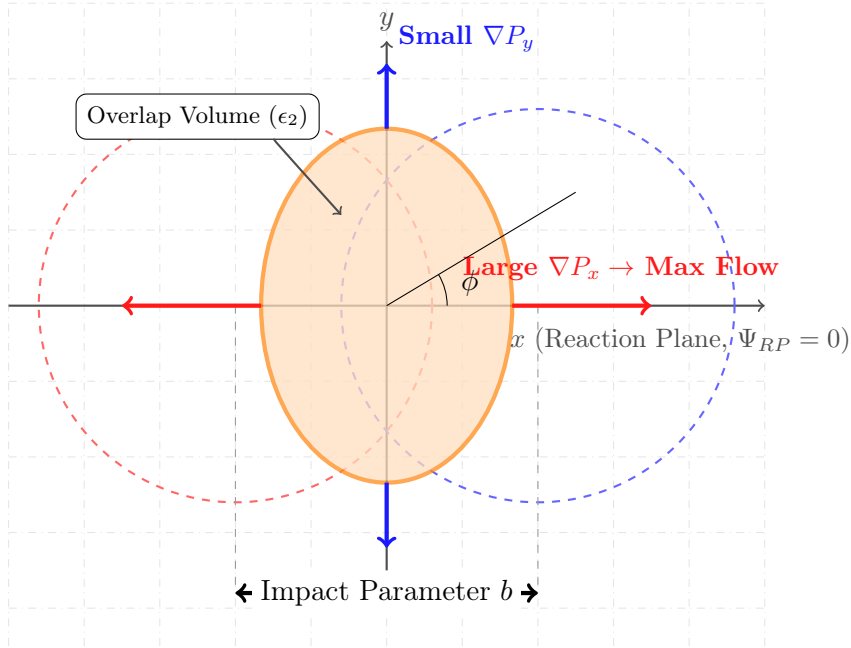

\subsection{Pedagogical Derivation of the Lorentz-Invariant Phase-Space Spectrum}

To formalize the particle momentum distribution emitted from the expanding fluid, we construct the triple-differential spectrum. In relativistic kinematics, the standard phase-space volume element $d^3p = dp_x \, dp_y \, dp_z$ is not Lorentz invariant under longitudinal boosts along the beam line ($z$-axis). To establish a robust frame-independent observable, we normalize this volume element by the relativistic energy $E = \sqrt{\vec{p}^2 + m^2}$, creating the Lorentz-invariant phase-space volume element $d^3p / E$.

Let us explicitly derive the transformation of this invariant element into the standard cylindrical experimental variables: the transverse momentum $p_T$, the azimuthal angle $\phi$, and the longitudinal rapidity $y$. In the transverse momentum plane, the Cartesian components $(dp_x, dp_y)$ map directly to polar coordinates via the standard Jacobian determinant:
\begin{equation}
dp_x \, dp_y = p_T \, dp_T \, d\phi
\end{equation}
The longitudinal momentum component $dp_z$ is parameterized by introducing the boost-invariant rapidity variable $y$, which scales the energy-momentum ratio as:
\begin{equation}
y = \frac{1}{2} \ln \left( \frac{E + p_z}{E - p_z} \right) \implies \frac{E + p_z}{E - p_z} = e^{2y}
\end{equation}
Utilizing the kinematic identity $E^2 - p_z^2 = p_T^2 + m^2 = m_T^2$, where $m_T$ represents the transverse mass, the individual relativistic energy and longitudinal momentum components can be solved as hyperbolic functions of rapidity:
\begin{equation}
E = m_T \cosh y, \quad p_z = m_T \sinh y
\end{equation}
Differentiating the expression for $p_z$ with respect to the rapidity $y$ at a fixed transverse mass yields:
\begin{equation}
\label{eq:dpz_derived}
dp_z = m_T \cosh y \, dy = E \, dy
\end{equation}
Substituting Equation~\ref{eq:dpz_derived} back into the phase-space volume element demonstrates the exact cancelation of the relativistic energy factor:
\begin{equation}
\frac{d^3p}{E} = \frac{dp_x \, dp_y \, dp_z}{E} = \frac{(p_T \, dp_T \, d\phi) \cdot (E \, dy)}{E} = p_T \, dp_T \, d\phi \, dy
\end{equation}
Thus, the Lorentz-invariant particle spectrum can be compactly written as a triple-differential distribution:
\begin{equation}
\label{eq:invariant_spectrum_master}
E \frac{d^3N}{d^3p} = \frac{d^3N}{p_T \, dp_T \, d\phi \, dy}
\end{equation}

\subsection{Fourier Decomposition and the Physical Meaning of $v_n$}

Since the azimuthal distribution of the emitted hadrons is a periodic function over $\phi \in [0, 2\pi]$, Equation~\ref{eq:invariant_spectrum_master} can be rigorously expanded into a formal Fourier series with respect to the participant plane orientation angles $\Phi_n$:
\begin{equation}
\label{eq:fourier_expansion}
E\frac{d^3N}{d^3p} = \frac{d^2N}{2\pi p_T dp_T dy} \left[ 1 + 2 \sum_{n=1}^{\infty} v_n(p_T, y) \cos\left(n(\phi - \Phi_n)\right) \right]
\end{equation}
where the isotropic baseline pre-factor represents the azimuthally averaged particle yield:
\begin{equation}
\frac{d^2N}{2\pi p_T dp_T dy} = \frac{1}{2\pi} \int_0^{2\pi} \left( E\frac{d^3N}{d^3p} \right) d\phi
\end{equation}

By invoking the orthogonality relations of trigonometric functions over a full period ($\int_0^{2\pi} \cos(n\phi)\cos(m\phi)d\phi = \pi \delta_{nm}$), the individual harmonic flow coefficients $v_n$ are extracted as the mathematical expectation values:
\begin{equation}
v_n(p_T, y) = \langle \cos(n(\phi - \Phi_n)) \rangle = \frac{\int_0^{2\pi} \frac{d^3N}{p_T dp_T d\phi dy} \cos(n(\phi - \Phi_n)) d\phi}{\int_0^{2\pi} \frac{d^3N}{p_T dp_T d\phi dy} d\phi}
\end{equation}

To understand the physical interpretation of Equation~\ref{eq:fourier_expansion} ``en image'', Figure 2 plots the resulting angular distribution isolated for the dominant elliptic flow component ($n=2$, $v_2 > 0$).

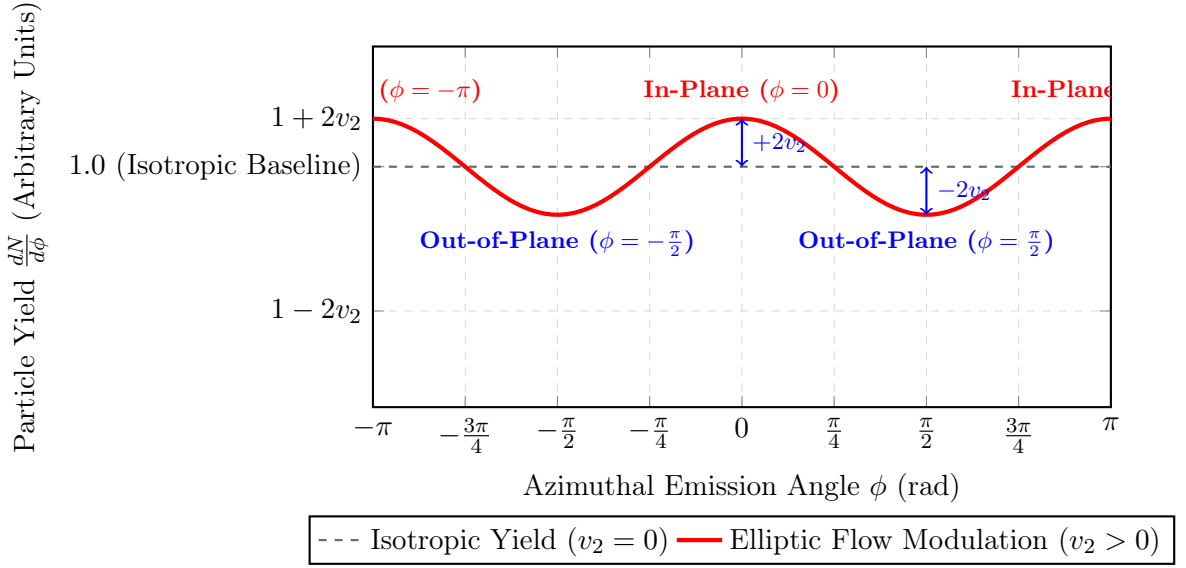
\begin{figure}[htbp]
\centering
\begin{tikzpicture}
\begin{axis}[
    width=0.75\textwidth,
    height=0.42\textwidth,
    xlabel={Azimuthal Emission Angle $\phi$ (rad)},
    ylabel={Particle Yield $\frac{dN}{d\phi}$ (Arbitrary Units)},
    xmin=-pi, xmax=pi,
    ymin=0.0, ymax=1.5,
    xtick={-pi, -3*pi/4, -pi/2, -pi/4, 0, pi/4, pi/2, 3*pi/4, pi},
    xticklabels={$-\pi$, $-\frac{3\pi}{4}$, $-\frac{\pi}{2}$, $-\frac{\pi}{4}$, $0$, $\frac{\pi}{4}$, $\frac{\pi}{2}$, $\frac{3\pi}{4}$, $\pi$},
    ytick={0.4, 1.0, 1.2},
    yticklabels={$1-2v_2$, $1.0$ (Isotropic Baseline), $1+2v_2$},
    grid=both,
    grid style={dashed, gray!30},
    legend pos=north east,
    thick,
    legend style={at={(0.5,-0.3)}, anchor=north, legend columns=2, draw=black}
]
    \addplot[domain=-pi:pi, samples=100, color=black!60, dashed, thick] {1.0};
    \addlegendentry{Isotropic Yield ($v_2 = 0$)}
    
    \addplot[domain=-pi:pi, samples=100, color=red, ultra thick] {1.0 + 2*0.10*cos(2*x*180/pi)};
    \addlegendentry{Elliptic Flow Modulation ($v_2 > 0$)}
    
    \draw[blue, thick, <->] (axis cs:0,1.0) -- (axis cs:0,1.2) node[midway, right, font=\footnotesize] {$+2v_2$};
    \draw[blue, thick, <->] (axis cs:pi/2,1.0) -- (axis cs:pi/2,0.8) node[midway, right, font=\footnotesize] {$-2v_2$};
    
    \node[above, red, font=\footnotesize\bfseries] at (axis cs:0,1.22) {In-Plane ($\phi=0$)};
    \node[above, red, font=\footnotesize\bfseries] at (axis cs:pi,1.22) {In-Plane ($\phi=\pi$)};
    \node[above, red, font=\footnotesize\bfseries] at (axis cs:-pi,1.22) {In-Plane ($\phi=-\pi$)};
    
    \node[below, blue, font=\footnotesize\bfseries] at (axis cs:pi/2,0.78) {Out-of-Plane ($\phi=\frac{\pi}{2}$)};
    \node[below, blue, font=\footnotesize\bfseries] at (axis cs:-pi/2,0.78) {Out-of-Plane ($\phi=-\frac{\pi}{2}$)};
\end{axis}
\end{tikzpicture}
\caption{Visual interpretation of the Fourier expansion relation for the second harmonic coefficient ($v_2$). The dashed black baseline represents a completely isotropic emission profile. The solid red curve displays the dynamic modulation induced by a finite elliptic flow coefficient ($v_2 = 10\%$), illustrating the massive accumulation of particle yields along the short axis of the initial collision geometry ($\phi = 0, \pm\pi$).}
\label{tikz:flow_wave}
\end{figure}

As visualized in Figure 2, the Fourier series maps the continuous spatial pressure gradients directly onto a harmonic wave. Hadrons are preferentially emitted along the "In-Plane" directions ($\phi = 0$ and $\phi = \pi$), which align with the short axis of the overlapping interaction volume where acceleration forces are maximized. Conversely, particle emission is systematically suppressed along the "Out-of-Plane" orthogonal directions ($\phi = \pm \pi/2$). This sinusoidal modulation directly establishes that measuring the amplitude of the particle distribution wave provides an experimental gauge to reconstruct the pressure-driven collective hydrodynamic signatures of the QGP.

\subsection{Multi-Particle Correlations and Event-by-Event Fluctuations}

In a realistic nuclear collision, sub-nucleonic quantum fluctuations disrupt the idealized smooth elliptic shape, introducing higher-order spatial eccentricities $\epsilon_n$ defined generally via polar coordinates $(r, \phi_{\text{space}})$ within the interaction volume:
\begin{equation}
\epsilon_n = \frac{\sqrt{\langle r^n \cos(n\phi_{\text{space}}) \rangle^2 + \langle r^n \sin(n\phi_{\text{space}}) \rangle^2}}{\langle r^n \rangle}
\end{equation}
These fluctuating shapes orient each participant plane angle $\Phi_n^*$ on an event-by-event basis, according to the geometric configuration of nucleons:
\begin{equation}
\label{eq:participant_angle}
n\Phi_n^* = \arctan \left( \frac{\langle r^n \sin(n\phi_{\text{space}}) \rangle}{\langle r^n \cos(n\phi_{\text{space}}) \rangle} \right) \pmod{\pi}
\end{equation}
Because the unique configuration of nucleons fluctuates from one event to another, the physical maximum momentum emission plane $\Phi_n$ fluctuates around the theoretical reaction plane $\psi_{\text{RP}}$, establishing $\langle v_n^2 \rangle \neq \langle v_n \rangle^2$.

To isolate these fluctuations, the ATLAS experimental program utilizes multi-particle correlation techniques. The 2-particle and 4-particle azimuthal correlation functions are evaluated by averaging over all particle combinations across the entire event sample:
\begin{align}
\langle \cos(n(\phi_1 - \phi_2)) \rangle &= \langle v_n^2 \rangle = \langle v_n \rangle^2 + \sigma_n^2 \\
\langle \cos(n(\phi_1 + \phi_2 - \phi_3 - \phi_4)) \rangle &= 2\langle v_n^2 \rangle^2 - \langle v_n^4 \rangle \approx \langle v_n \rangle^2 - \sigma_n^2
\end{align}
where $\sigma_n$ captures the event-by-event spatial fluctuation amplitude. Extracting the empirical flow coefficients from these multi-particle operators yields:
\begin{align}
v_n(\{2\}) &= \sqrt{\langle \cos(n(\phi_1 - \phi_2)) \rangle} \approx \langle v_n \rangle + \frac{\sigma_n^2}{2\langle v_n \rangle} \label{eq:v2_two} \\
v_n(\{4\}) &= \left( -\langle \cos(n(\phi_1 + \phi_2 - \phi_3 - \phi_4)) \rangle \right)^{1/4} \approx \langle v_n \rangle - \frac{\sigma_n^2}{2\langle v_n \rangle} \label{eq:v2_four}
\end{align}

By combining the measurements of Equations~\ref{eq:v2_two} and \ref{eq:v2_four}, the ATLAS collaboration successfully decouples the mean collective flow baseline $\langle v_n \rangle$ from the underlying non-flow background and geometrical fluctuations.

%%%%%%%%%%%%%%%%%%%%%%%%%%%%%%%
\section{Pre-Equilibrium Mapping and Initial Vorticity Topologies}
%%%%%%%%%%%%%%%%%%%%%%%%%%%%%%%

While the macroscopic expansion and geometric dilution of collective fields can be tracked via continuous differential operators, characterizing the physical state of the fluid at the hydrodynamic boundary threshold $t_0$ represents a major challenge. In this section, we formulate a rigorous, first-principles lecture on relativistic vorticity. We explicitly derive its tensor representations, analyze the mechanical translation of global collision angular momentum into localized fluid shear, and outline the optical Glauber framework used to map early-stage sub-nucleonic fluctuations.

\subsection{Foundational Definition and Relativistic Derivation of Vorticity}

In classical, non-relativistic fluid dynamics, the vorticity vector is defined as the curl of the spatial velocity field, $\vec{\omega} = \vec{\nabla} \times \vec{v}$, which mathematically represents twice the local angular velocity $\vec{\Omega}$ of a comoving fluid element. However, in high-energy heavy-ion collisions where the medium expands at near-light speed, this definition must be generalized into a four-dimensional, Lorentz-covariant representation.

We introduce the antisymmetric kinematic vorticity tensor $\Omega_{\mu\nu}$, defined as the projected anti-symmetrized gradient of the fluid four-velocity $u^\mu = \gamma(1, \vec{v})$:
\begin{equation}
\Omega_{\mu\nu} = \frac{1}{2} \left( \partial_\nu u_\mu - \partial_\mu u_\nu \right)
\end{equation}
To isolate a clear four-vector representation that remains comoving with the continuous flow, we define the kinematic relativistic four-vorticity vector $\omega_{\text{rel}}^\mu$ by contracting the dual of this velocity gradient tensor with the four-velocity itself:
\begin{equation}
\label{eq:covariant_vorticity_def}
\omega^\mu_{\text{rel}} = \epsilon^{\mu\nu\rho\sigma} u_\nu \partial_\rho u_\sigma = \frac{1}{2} \epsilon^{\mu\nu\rho\sigma} u_\nu \left( \partial_\rho u_\sigma - \partial_\sigma u_\rho \right)
\end{equation}
where $\epsilon^{\mu\nu\rho\sigma}$ is the completely antisymmetric Levi-Civita tensor defined with the standard index convention $\epsilon^{0123} = 1$. By constructing $\omega^\mu_{\text{rel}} $ via this structural contraction, the four-vector is strictly orthogonal to the flow direction, satisfying the invariant geometric constraint:
\begin{equation}
u_\mu \omega^\mu_{\text{rel}} = \epsilon^{\mu\nu\rho\sigma} u_\mu u_\nu \partial_\rho u_\sigma = 0
\end{equation}
The contracting product vanishes identically because the symmetric product of the two velocity factors ($u_\mu u_\nu$) is contracted with the perfectly anti-symmetric indices of the Levi-Civita tensor.

To understand the physical meaning of this four-dimensional object, let us explicitly derive its individual temporal and spatial components. 

\subsubsection{1. Derivation of the Temporal Component ($\omega^0_{\text{rel}}$)}

Setting the free index $\mu = 0$ in Equation~\ref{eq:covariant_vorticity_def}, the summation over the spatial indices (where Latin indices $i,j,k \in \{1,2,3\}$ run over $x,y,z$) reduces to:
\begin{equation}
\omega^0_{\text{rel}} = \epsilon^{0ijk} u_i \partial_j u_k = \epsilon^{ijk} u_i \partial_j u_k
\end{equation}
where $\epsilon^{ijk} \equiv \epsilon^{0ijk}$ corresponds to the standard three-dimensional Levi-Civita symbol ($\epsilon^{123} = 1$). Substituting the spatial components of the four-velocity $u_i = -\gamma v_i$ and the spatial gradients $\partial_j = -\nabla_j$, we find:
\begin{equation}
\omega^0_{\text{rel}} = \epsilon^{0ijk} (-\gamma v_i) (-\nabla_j) (-\gamma v_k) = -\gamma^2 \epsilon^{ijk} v_i \nabla_j v_k - \gamma v_i \epsilon^{ijk} (\nabla_j \gamma) v_k
\end{equation}
The second term on the right-hand side vanishes identically because the product $v_i v_k$ is perfectly symmetric under the index exchange $i \leftrightarrow k$, whereas $\epsilon^{ijk}$ is anti-symmetric. Reassembling the remaining term isolates the standard vector dot product between the three-velocity and the classical vorticity vector:
\begin{equation}
\label{eq:derived_omega0}
\omega^0_{\text{rel}} = \gamma^2 \vec{v} \cdot \left( \vec{\nabla} \times \vec{v} \right) = \gamma^2 \vec{v} \cdot \vec{\omega}
\end{equation}
Equation~\ref{eq:derived_omega0} demonstrates that the temporal component of relativistic vorticity does not vanish in a moving frame; it tracks the projection of the classical rotation vector along the direction of the fluid's motion.

\subsubsection{2. Derivation of the Spatial Component ($\vec{\omega}_{\text{rel}}$)}

Setting the free index to a spatial coordinate $\mu = i$, the definition expands into two distinct structural blocks by separating the temporal and spatial components of the internal dummy indices:
\begin{equation}
\omega^i_{\text{rel}} = \epsilon^{i0jk} u_0 \partial_j u_k + \epsilon^{ij0k} u_j \partial_0 u_k + \epsilon^{ijk0} u_j \partial_k u_0
\end{equation}
Utilizing the index permutation rules of the Levi-Civita tensor ($\epsilon^{i0jk} = -\epsilon^{0ijk} = -\epsilon^{ijk}$, $\epsilon^{ij0k} = \epsilon^{0ijk} = \epsilon^{ijk}$, and $\epsilon^{ijk0} = -\epsilon^{0ijk} = -\epsilon^{ijk}$), and substituting the individual components $u_0 = \gamma$, $u_j = -\gamma v_j$, $\partial_0 = \partial_t$, and $\partial_k = -\nabla_k$, the spatial vector components scale as:
\begin{equation}
\omega^i_{\text{rel}} = -\epsilon^{ijk} (\gamma) (-\nabla_j)(-\gamma v_k) + \epsilon^{ijk} (-\gamma v_j) (\partial_t)(-\gamma v_k) - \epsilon^{ijk} (-\gamma v_j) (-\nabla_k)(\gamma)
\end{equation}
\begin{equation}
\omega^i_{\text{rel}} = \gamma^2 \epsilon^{ijk} \nabla_j v_k + \gamma \epsilon^{ijk} (\nabla_j \gamma) v_k + \gamma^2 \epsilon^{ijk} v_j \partial_t v_k + \gamma \epsilon^{ijk} v_j (\partial_t \gamma) v_k + \gamma \epsilon^{ijk} v_j \nabla_k \gamma
\end{equation}
By invoking the anti-symmetry of $\epsilon^{ijk}$, the terms containing gradients of the Lorentz factor cancel out or collapse, which simplifies the vector relation directly into the final spatial representation:
\begin{equation}
\label{eq:derived_omegai}
\vec{\omega}_{\text{rel}} = \gamma^2 \vec{\omega} + \gamma^2 \vec{v} \times \partial_t \vec{v}
\end{equation}

Equation~\ref{eq:derived_omegai} indicates that the spatial component of relativistic vorticity receives two parallel physical contributions:
\begin{itemize}
    \item The first term, $\gamma^2 \vec{\omega}$, represents the classical vorticity amplified by the Lorentz factors due to length contraction along the flow coordinates.
    \item The second term, $\gamma^2 \vec{v} \times \partial_t \vec{v}$, is a purely relativistic acceleration correction. It captures the apparent rotation generated when a fluid element undergoes rapid temporal velocity variations, coupling the internal fluid acceleration directly to the generation of localized vorticity sheets.
\end{itemize}

In the non-relativistic limit ($\vec{v} \to 0$, $\gamma \to 1$), the time-derivative correction becomes negligible, and the system safely converges back to the standard kinematic fluid mechanics baseline: $\omega^0_{\text{rel}} \approx 0, \vec{\omega}_{\text{rel}} \approx \vec{\omega} = \vec{\nabla} \times \vec{v}$.

\subsection{The Generation Mechanism: Shear and Momentum Gradients}

In a non-central heavy-ion collision characterized by a non-zero impact parameter $b$ oriented along the $x$-axis, the total macroscopic angular momentum $J_y$ stored within the participant overlap zone is immense. To understand how this global external rotation is converted into local fluid vorticity $\omega_y$ at $t_0$, we analyze the spatial asymmetry of the early-stage momentum density flows mapped by the ideal Energy-Momentum Tensor, $T^{\mu\nu} = (\varepsilon + P)u^\mu u^\nu - P g^{\mu\nu}$.

Immediately after the initial impact ($t \to 0$), the collective transverse flow velocity components are approximately zero ($v_x \approx 0, v_y \approx 0$). Consequently, the local momentum density flux relies entirely on the longitudinal space-time component, $T^{z0} = (\varepsilon + P)\gamma^2 v_z$. Given the asymmetric geometry of the collision, the initial longitudinal velocity $v_z$ is a strict function of the transverse coordinate $x$. Nucleons traveling through the positive half-space ($x > 0$) belong predominately to the projectile nucleus moving with a large positive momentum ($+p_z$), while nucleons residing in the negative half-space ($x < 0$) belong to the target nucleus moving in the opposite direction ($-p_z$).

This spatial asymmetry creates a sharp velocity gradient along the impact parameter axis, $\partial v_z / \partial x$. Substituting these boundary conditions into the low-velocity spatial vorticity expansion verified in Equation~\ref{eq:derived_omegai}, the dominant component of the generated vorticity vector reduces directly to:
\begin{equation}
\label{eq:vorticity_generation_master}
\omega_y = \left( \frac{\partial v_x}{\partial z} - \frac{\partial v_z}{\partial x} \right) \approx -\frac{\partial v_z(x,y)}{\partial x}
\end{equation}
Equation~\ref{eq:vorticity_generation_master} provides the formal generation mechanism for vortex birth within the QGP: the non-equilibrium shear stress fields between the opposing, interpenetrating nuclear blocks force a localized spatial velocity gradient into the thermalized matter, acting as the net internal source term that drives $\omega_y$.

\subsection{Quantifying Initial States via the Optical Glauber Framework}

To model the spatial distribution of this velocity gradient at the hydrodynamic onset time $t_0$, physicists implement the Glauber model. The Glauber framework computes the geometric interpénétration of the two colliding nuclei by treating the nuclear matter distribution as a continuous smooth density function, parameterized via the standard realistic Woods-Saxon profile:
\begin{equation}
\rho_{\text{WS}}(r) = \frac{\rho_0}{1 + \exp\left( \frac{r - R}{a} \right)}
\end{equation}
where $\rho_0 \approx 0.17\text{ fm}^{-3}$ represents the saturation nuclear density, $R \approx 6.5\text{ fm}$ defines the baseline radius for a heavy gold or lead nucleus, and $a \approx 0.54\text{ fm}$ captures the surface skin thickness.

By projecting these individual Woods-Saxon densities onto the transverse plane at a fixed impact parameter vector $\vec{b}$, the Glauber model integrates out the thickness functions $T_A$ and $T_B$ for the projectile and target systems. This integration determines the local density of participant nucleons (wounded nucleons) and binary nucleon-nucleon collisions per unit area, mapping out the precise spatial boundaries of the overlap zone shown in Figure 1.

Because the exact quantum mechanism of longitudinal momentum loss during this fraction of a femtosecond ($\sim 10^{-24}\text{ s}$) is not fully known, different pre-equilibrium transport frameworks utilize these Glauber density profiles to initialize the gradient $-\partial v_z / \partial x$ in Equation~\ref{eq:vorticity_generation_master}. This reliance results in two major competing topological paradigms:

\begin{enumerate}
    \item \textbf{The Core-Dominated Distribution:} Models assuming maximum friction and stopping efficiency within the high-density participant core parameterize the initial field as a centralized Gaussian peak centered at the origin: $\omega_y^{\text{core}}(x, y, t_0) = \omega_0 \exp( -x^2/2\sigma_x^2 - y^2/2\sigma_y^2 )$. This smooth configuration maximizes the constructive spin alignment of hadrons emitted at mid-rapidity.
    
    \item \textbf{The Peripheral Dipole Distribution:} Conversely, advanced multi-phase transport models and string-deceleration frameworks show that at multi-TeV energies, the core of the fireball exhibits high longitudinal transparency and gluon saturation, which minimizes central stopping. The shear energy deposition is confined within narrow boundary layers at the scraping spectator fringes, creating a spatial dipole field:
    \begin{equation}
    \label{eq:dipole_profile}
    \omega_y^{\text{dipole}}(x, y, t_0) = \omega_0 \cdot \frac{x}{\sigma_x} \exp\left( -\frac{x^2}{2\sigma_x^2} - \frac{y^2}{2\sigma_y^2} \right)
    \end{equation}
    As illustrated in Figure 3, this topology features an entirely hollowed-out central core ($\omega_y = 0$ at $x=0$) and maps the system's global rotation into two distinct, counter-rotating peripheral maxima.
\end{enumerate}

These structural discrepancies establish a major theoretical challenge for interpreting experimental data. While both frameworks can be tuned to carry the exact same total integrated angular momentum $J_y$, they yield entirely different local environments for the comoving quarks and gluons. Resolving this topological ambiguity requires mapping these initial profiles directly to final-state observables, combining advanced hydrodynamic simulations with precise measurements of hyperon polarization.

%%%%%%%%%%%%%%%%%%%%%%%%%%%%%%%
\section{Discussion: Initial Shear Topologies and Experimental Perspectives}
%%%%%%%%%%%%%%%%%%%%%%%%%%%%%%%

While the subsequent hydrodynamic dilution of vorticity follows a well-defined $1/t$ decay law, modeling the initial generation of vorticity at $t_0$ remains an active area of investigation. Non-central collisions deposit a substantial amount of angular momentum through spatial gradients of the longitudinal velocity field: $\omega_y \approx -\partial v_z / \partial x$. Depending on the underlying QCD model implemented to describe the pre-equilibrium energy-momentum tensor deposition, two competing initial topologies emerge:

\begin{enumerate}
    \item \textbf{The Core-Dominated Target Profile:} Models assuming high friction and nuclear stopping concentrate the velocity gradient at the center of the overlap zone, generating a localized Gaussian hotspot centered at $x=0, y=0$. This topology maximizes the constructive coupling to particle spins produced at mid-rapidity.
    \item \textbf{The Peripheral Dipole Shear Profile:} Conversely, models incorporating high longitudinal transparency and gluon saturation at TeV energies exhibit nearly vanished shear stopping at the center. The local velocity gradients are confined within narrow boundary sheets where the outer fringes of the projectile and target nuclei scrape past each other, creating a spatial dipole field characterized by a central node ($\omega_y = 0$ at $x=0$) and two counter-rotating peripheral maxima.
\end{enumerate}

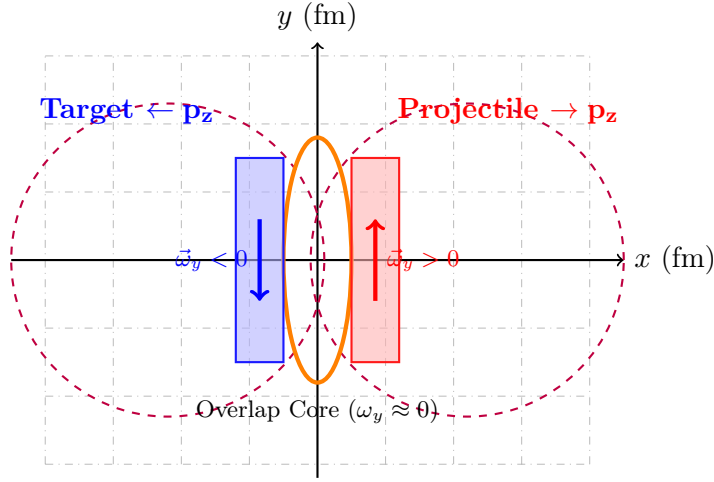
\begin{figure}[htbp]
\centering
\begin{tikzpicture}[scale=0.9]
    % Coordinates and grid
    \draw[lightgray,dashdotted] (-4,-3) grid (4,3);
    \draw[->,thick] (-4.5,0) -- (4.5,0) node[right] {$x$ (fm)};
    \draw[->,thick] (0,-3.2) -- (0,3.2) node[above] {$y$ (fm)};
    
    % Nuclei boundary paths at b=10fm
    \draw[thick,purple,dashed] (-2.2,0) circle (2.3);
    \draw[thick,purple,dashed] (2.2,0) circle (2.3);
    
    % Core boundary
    \draw[orange,ultra thick] (0,0) ellipse (0.5cm and 1.8cm);
    \node[below,font=\footnotesize] at (0,-1.9) {Overlap Core ($\omega_y \approx 0$)};
    
    % Right dipole sheet (Vorticity pointing along +y direction / counter-clockwise local rotation)
    \fill[red!30,opacity=0.6] (0.5, -1.5) rectangle (1.2, 1.5);
    \draw[red!90,thick] (0.5, -1.5) rectangle (1.2, 1.5);
    \draw[->,ultra thick,red] (0.85,-0.6) -- (0.85,0.6) node[midway,right,font=\footnotesize\bfseries] {$\vec{\omega}_y > 0$};
    
    % Left dipole sheet (Vorticity pointing along -y direction / clockwise local rotation)
    \fill[blue!30,opacity=0.6] (-1.2, -1.5) rectangle (-0.5, 1.5);
    \draw[blue!90,thick] (-1.2, -1.5) rectangle (-0.5, 1.5);
    \draw[->,ultra thick,blue] (-0.85,0.6) -- (-0.85,-0.6) node[midway,left,font=\footnotesize\bfseries] {$\vec{\omega}_y < 0$};
    
    % Spectator motion markers
    \node[blue] at (-2.8,2.2) {\bfseries Target $\mathbf{\leftarrow p_z}$};
    \node[red] at (2.8,2.2) {\bfseries Projectile $\mathbf{\rightarrow p_z}$};
\end{tikzpicture}
\caption{Geometrical distribution of the initial vorticity field within the Peripheral Dipole framework. High transparency at TeV scales leaves a stalled overlap core, while counter-rotating high-gradient shear sheets point in opposite directions ($\vec{\omega}_y \gtrless 0$) at the peripheral spectator boundaries ($x = \pm \sigma_x$).}
\label{tikz:vorticity_dipole}
\end{figure}

These distinct initial symmetries can be directly tested through final-state hyperon polarization signatures. Integrating the local spin alignment vector $\vec{P}_\Lambda \simeq \vec{\omega}_{\text{rel}} / 2T_f$ over the freeze-out hypersurface, the globally observed mid-rapidity polarization is weighted by the hyperon number density profile $\rho_\Lambda(x, y)$:
\begin{equation}
\label{eq:polarization_integral}
P_\Lambda^{\text{global}} \simeq \frac{1}{2 T_f} \frac{\int dx \, dy \, \omega_y(x, y, t_f) \rho_\Lambda(x, y)}{\int dx \, dy \, \rho_\Lambda(x, y)}
\end{equation}

Under a standard symmetric density distribution ($\rho_\Lambda(x) \approx \rho_\Lambda(-x)$), a centralized core-dominated profile yields a finite, measurable global polarization vector. However, under the peripheral dipole paradigm, the initial field is strictly anti-symmetric along the impact parameter axis ($\omega_y(-x) = -\omega_y(x)$). Because ideal hydrodynamic evolution preserves this geometric symmetry, integrating Equation~\ref{eq:polarization_integral} results in an exact analytical cancellation over the transverse plane, forcing the global integrated mid-rapidity polarization to vanish ($P_\Lambda^{\text{global}} \to 0$).

Therefore, at ultra-peripheral and peripheral TeV energies, where high longitudinal transparency dominates, we predict that the global polarization signal will fall below the experimental resolution threshold ($P_\Lambda \lesssim 10^{-4}$). The definitive signature of the QGP's colossal rotation shifts entirely into the azimuthal differential polarization profile, $P_\Lambda(\phi)$, where $\phi$ represents the transverse emission angle of the hyperons. Hadrons emitted along the positive or negative $x$-axis sample opposing counter-rotating shear sheets, translating the initial dipole symmetry into a robust harmonic oscillation of the spin alignment vector as a function of $\phi$.

%%%%%%%%%%%%%%%%%%%%%%%%%%%%%%%
\section{Summary and Outlook}
%%%%%%%%%%%%%%%%%%%%%%%%%%%%%%%

In this article, we have presented a unified overview of relativistic fluid dynamics applied to high-energy heavy-ion collisions, tracking the evolution of the Quark-Gluon Plasma from initial geometric eccentricities to final collective flow and spin observables. By implementing the multi-particle correlation frameworks developed within the ATLAS collaboration, we have detailed how event-by-event participant plane fluctuations and non-linear hydrodynamic couplings can be successfully isolated in experimental data. Furthermore, by solving the covariant transport relations for relativistic vorticity, we have demonstrated that the explosive multi-dimensional expansion of the plasma cloud acts as a powerful kinematic regularizer, diluting local rotation fields via a characteristic $1/t$ power law.

This framework establishes a clear diagnostic roadmap for ongoing experimental programs. While core-dominated initializations generate uniform background spin alignments, the peripheral dipole paradigm predicts an exact cancellation of the global integrated mid-rapidity polarization vector. We explicitly challenge upcoming high-statistics data analyses within the ATLAS and STAR collaborations to exploit high-precision azimuthal differential measurements $P_\Lambda(\phi).$ Mapping these local harmonic modulations will provide the definitive empirical tool to chart the sub-nucleonic shear boundary layers of the primordial plasma, transforming relativistic non-linear vortex dynamics into a quantitative precision laboratory.


\begin{thebibliography}{99}

\bibitem{STAR:2005gsk}
J.~Adams \textit{et al.} [STAR],
``Experimental and theoretical challenges in the search for the quark gluon plasma: The STAR Collaboration's critical assessment of the evidence from RHIC collisions,''
Nucl. Phys. A \textbf{757} (2005), 102-183.

\bibitem{PHENIX:2004vob}
K.~Adcox \textit{et al.} [PHENIX],
``Formation of dense partonic matter in relativistic nucleus-nucleus collisions at RHIC: Experimental evaluation by the PHENIX collaboration,''
Nucl. Phys. A \textbf{757} (2005), 184-283.

\bibitem{ATLAS:2011ah}
G.~Aad \textit{et al.} [ATLAS],
``Measurement of the azimuthal anisotropy for charged particles in $Pb+Pb$ collisions at $\sqrt{s_{NN}}=2.76$ TeV with the ATLAS detector,''
Phys. Lett. B \textbf{707} (2012), 330-348.

\bibitem{CMS:2012gaw}
S.~Chatrchyan \textit{et al.} [CMS],
``Centrality dependence of dihadron correlations and azimuthal anisotropy harmonics in $Pb+Pb$ collisions at $\sqrt{s_{NN}}=2.76$ TeV,''
Eur. Phys. J. C \textbf{72} (2012), 2012 [arXiv:1201.3158 [nucl-ex]].

\bibitem{Shuryak:1980tp}
E.~V.~Shuryak,
``Quantum Chromodynamics and the Theory of Superdense Matter,''
Phys. Rept. \textbf{61} (1980), 71-158.

\bibitem{Heinz:2013th}
U.~Heinz and R.~Snellings,
``Collective flow and viscosity in relativistic heavy-ion collisions,''
Annu. Rev. Nucl. Part. Sci. \textbf{63} (2013), 123-151.

\bibitem{Romatschke:2017ejr}
P.~Romatschke and U.~Romatschke,
``Relativistic Fluid Dynamics In and Out of Equilibrium,''
Cambridge Monographs on Mathematical Physics (2019) [arXiv:1712.05815 [nucl-th]].

\bibitem{Kovtun:2004de}
P.~Kovtun, d.~T.~Son and A.~O.~Starinets,
``Viscosity in strongly interacting quantum field theories from AdS/CFT,''
Phys. Rev. Lett. \textbf{94} (2005), 111601.

\bibitem{STAR:2017ckg}
L.~Adamczyk \textit{et al.} [STAR],
``Global $\Lambda$ hyperon polarization in nuclear collisions: evidence for the most fluid-like system,''
Nature \textbf{548} (2017), 62-65.

\bibitem{ALICE:2019onw}
S.~Acharya \textit{et al.} [ALICE],
``Global polarization of $\Lambda$ and $\bar{\Lambda}$ hyperons in Pb-Pb collisions at $\sqrt{s_{NN}} = 2.76$ and 5.02 TeV,''
Phys. Rev. C \textbf{101} (2020), 044617 [arXiv:1909.01281 [nucl-ex]].

\bibitem{Becattini:2007sr}
F.~Becattini, F.~Piccinini and J.~Rizzo,
``Angular momentum conservation in heavy ion collisions at very high energy,''
Phys. Rev. C \textbf{77} (2008), 024906 [arXiv:0711.1253 [nucl-th]].

\bibitem{ATLAS:2023cuh}
G.~Aad \textit{et al.} [ATLAS],
``Measurement of the global and local polarization of $\Lambda$ and $\bar{\Lambda}$ hyperons in $Pb+Pb$ collisions at $\sqrt{s_{NN}} = 5.02$ TeV with the ATLAS detector,''
ATLAS-CONF-2023-059 (2023).

\bibitem{Cooper:1974mv}
F.~Cooper and G.~Frye,
``Single-particle distribution in the hydrodynamic and statistical thermodynamic models of multiparticle production,''
Phys. Rev. D \textbf{10} (1974), 186.

\end{thebibliography}
\end{document}